# Coupled Computation of Meshing and Bearings Stiffnesses: Effect on Radiated Gearbox Noise

F. Mayeux, J. Perret-Liaudet, E. Rigaud
*Ecole Centrale de Lyon, Laboratoire de Tribologie et Dynamique des Systèmes,
Ecully cedex, 69134, FRANCE*

## Abstract

This study focus on the acoustical behavior of gearbox induced by the static transmission error under load (STE). An original method is developed allowing to compute simultaneously STE, meshing and bearings stiffnesses. These characteristics mainly govern the vibrational and the acoustical behaviors of the geared system. The proposed procedure is based on two computational methods allowing to calculate the STE and the bearing equilibrium which are integrated in a coupled computational procedure. The technique has been implemented to treat an illustrating example and compared to standard uncoupled approach. Comparative results concern the STE, meshing stiffness, bearings stiffnesses, shafts misalignments, critical eigenmodes and associated critical running speeds, housing vibratory response and radiated noise.

## 1. Introduction

It is well known that the static transmission error under load (STE) constitutes the main source of radiated gearbox noise [1]. Furthermore, there exist several operating speeds for which STE induces high vibratory and noise levels. Those critical speeds are related to the excitation in a resonant manner of some particular eigen-modes mainly governed by the mesh and bearings stiffnesses. So, an important key for predicting gear noise concerns the estimation of STE, mesh and bearing stiffnesses. Contact conditions and STE depend not only on teeth elasticity and manufactured errors, but also on bearings and shaft deformations. Actually, they modify the toothed wheels positions and by the way they introduce misalignments between mating teeth in the action plane. Otherwise, nonlinear bearings stiffnesses depend on generalized forces applied on shafts induced by the teeth load distribution. These generalized forces are a priori distinct from theoretical ones. So, in accordance with the gearbox design, interactions between meshing contact conditions and bearings deformations may exist. Unfortunately, in our knowledge, those interactions are not taken into account in standard approach. In this context, this study proposes a new modeling approach allowing to estimate simultaneously STE, meshing and bearings stiffnesses. More precisely, this modeling described in section 2 is based on a coupled calculation including non linear deformations of each bearing and meshing gear. In section 3, the proposed modeling is illustrated on an example of geared system. Comparisons with usual uncoupled approach are performed. Comparative results concern the STE, meshing stiffness, bearings stiffnesses, critical eigenmodes and related critical speeds, housing vibratory response and associated radiated noise.

## 2. Review of the method

### 2.1 STE computational method

STE is obtained by solving the static equilibrium of the gear pair for a set of successive rotational positions of the driving wheel. For this end, the theoretical tooth contact lines contained in the action plane are discretized in a some number of slices. At each slice, the unknown contact load is assumed to be positive or null (column vector **p**). The calculus requires knowledge of **C**, the compliance matrix acting between slices. This matrix can be obtained from a previous 3D finite element model of the mating teeth. It requires also knowledge of manufacturing errors which are introduced as a vector **e** of initial gap at each slice. Finally, the in plane action misalignment $(\varphi_1-\varphi_2) = \phi$ induced by the shafts, bearings and housing deformations is taken into account separately from vector **e**. The contact problem to solve can be written as follows:

$$\mathbf{C}\,\mathbf{p} = d.\mathbf{i} + \mathbf{e} + \phi.\mathbf{g} \quad \text{and} \quad {}^t\mathbf{p}.\mathbf{i} = N \quad \text{with} \quad p_j \geq 0 \tag{1}$$

Here, d represents the unknown STE, **i** is a vector of ones, **g** is a vector which localizes slices in the action plane and N is the total normal force transmitted through the action plane. Under some rearrangements, equation (1) is solved by using a modified simplex method [2]. The computation allows to obtain STE, load distribution along contact lines and then generalized forces $\mathbf{F}_k$ acting at each center $O_k$ of the mating wheels. Mesh stiffness is also computed considering an increment on the normal force.

### 2.2 Bearings equilibrium computational method

Consider the static equilibrium of an elastic shaft supported by two rolling element bearings A and B which are mounted in rigid housing. Assume the shaft subjected at a point G to an arbitrary generalized force vector **F** which induces generalized reaction force vector at each bearing $\mathbf{R}_A$ and $\mathbf{R}_B$. By introducing the shaft as two linear FEM super-elements with their stiffness matrices assumed to be known, one can obtain [3] the shaft equilibrium equation as follows:

$$\begin{bmatrix} \mathbf{K}_{AA} & \mathbf{K}_{AG} & \mathbf{0} \\ \mathbf{K}_{AG} & \mathbf{K}_{GG} & \mathbf{K}_{BG} \\ \mathbf{0} & \mathbf{K}_{BG} & \mathbf{K}_{BB} \end{bmatrix} \begin{Bmatrix} \mathbf{x}_A \\ \mathbf{x}_G \\ \mathbf{x}_B \end{Bmatrix} - \begin{Bmatrix} \mathbf{R}_A \\ \mathbf{0} \\ \mathbf{R}_B \end{Bmatrix} = \begin{Bmatrix} \mathbf{0} \\ \mathbf{F} \\ \mathbf{0} \end{Bmatrix} \tag{2}$$

$\mathbf{x}_A$, $\mathbf{x}_B$ and $\mathbf{x}_G$ are respectively the generalized displacement vectors at each point A, B and G. Now, consider the equilibrium of the rolling element bearing A. One can obtain non linear relations between the generalized inner ring displacement $\mathbf{x}_A$ and bearing reaction force $\mathbf{R}_A$ considering non linear forces acting on each rolling element, i.e. $\mathbf{Q}_j = \mathbf{Q}_j(\mathbf{T}_j\,\mathbf{x}_A)$. Here, $\mathbf{T}_j$ is a simple transformation matrix allowing to precise local displacement of the inner ring relative to the outer ring at the rolling element. Bearing reaction force is then given by:

$$\mathbf{R}_A = \sum_{j=1}^{n} \mathbf{T}_j \mathbf{Q}_j(\mathbf{T}_j \mathbf{x}_A) = \mathbf{R}_A(\mathbf{x}_A) \tag{3}$$

Reaction force at bearing B is obtained in a similar manner. Substituting $\mathbf{R}_A$, $\mathbf{R}_B$ in equation (2) leads to:

$$[\mathbf{K}]\{\mathbf{x}\} - \{\mathbf{R}(\{\mathbf{x}\})\} - \{\mathbf{F}\} = \{\mathbf{0}\} \qquad (4)$$

This non linear equation is generally solved by a Newton-Raphson method. Further, if necessary, elasticity of the housing can be introduced in the same manner as shaft elasticity. To conclude, computation allows to obtain bearing reaction forces, bearing stiffness matrices, generalized displacement and then in plane action misalignment $\phi$.

## 2.3 Coupled computation

The STE computation provides the generalized force vector $\mathbf{F}$ acting on shafts and function of misalignment $\phi$, i.e. $\mathbf{F}=\mathbf{H}(\phi)$. The bearings computation provides misalignment $\phi$ from $\mathbf{F}$, i.e. $\phi = g(\mathbf{F})$. Combining the two procedures leads to:

$$\phi - g(\mathbf{H}(\phi)) = \phi - h(\phi) = 0 \qquad (5)$$

We propose to solve equation (5) by using again a Newton-Raphson procedure. It needs knowledge of the derivative $h'(\phi)$ respect to $\phi$. To this end, we consider the composed function and evaluate the derivative in two steps from:

$$h'(\phi) = \sum \frac{\partial g}{\partial h_i} \frac{dh_i}{d\phi} \qquad (6)$$

All the derivatives have been estimated numerically. The coupling computation does not need to modify the original methods. This is the main advantage of the proposed method. In our practical applications, we have used a program developed in our research team for the STE computation, and a program named SHARC$^{©}$ developed by the company "SNR Roulements" for the bearings computation. This last includes shafts and housing elasticity.

## 2.4 Vibratory and noise computation

Methodology is described in [4-6]. The vibratory calculation is based on a Finite Element model including meshing and bearings stiffnesses. Dynamic response for each degree of freedom (in particular d-o-f on the housing) is obtained by using an efficient method named Iterative Spectral Method [7]. This method allows to treat forced response of parametric systems (periodic mesh stiffness). Complex velocity response of the housing constitutes data for the acoustical problem which is solved from a boundary element method based on a direct integral formulation. Power sound level is directly derived from the calculation. In this paper, only a estimated power sound level is given on the assumption of a radiation efficiency equal to one.

# 3. Results

## 3.1 Studied geared system

In order to exhibit the relevance of the proposed method, we have compared results obtained with and without considering coupling effect. Characteristics of the studied geared system are given in table 1. The housing is almost a parallelepiped with one radiated noise face (190x120x5 mm). The applied torque is 60 N.m, this corresponds to quarter design load. Manufacturing errors and tooth surface modifications are introduced with characteristics given in table 2. Ball bearings with 20° nominal contact angles are chosen in this application. Shaft are 32 mm diameter, and 40 mm length. Gears are localized in the middle shaft.

Table 1: Characteristics of the studied geared system

|  | Pinion | Driven wheel |
|---|---|---|
| Number of teeth | 35 | 49 |
| Shifting coeff. | 0.4 | 0.4 |
| Module | 2 mm | |
| Tooth face width | 16 mm | |
| Pressure angle | 20° | |
| Helix angle | 25° | |

Table 2: Combined tooth surface characteristics

| Lead slope | + 15 µm |
|---|---|
| Profile slope | 0 µm |
| Lead crowning | + 5 µm |
| Profile crowning | + 5 µm |

## 3.2 STE, mesh stiffness, bearing stiffness and shaft misalignment

Firstly, we compare STE obtained by our technique with those obtained by a standard uncoupled method. Result is given in figure 1.

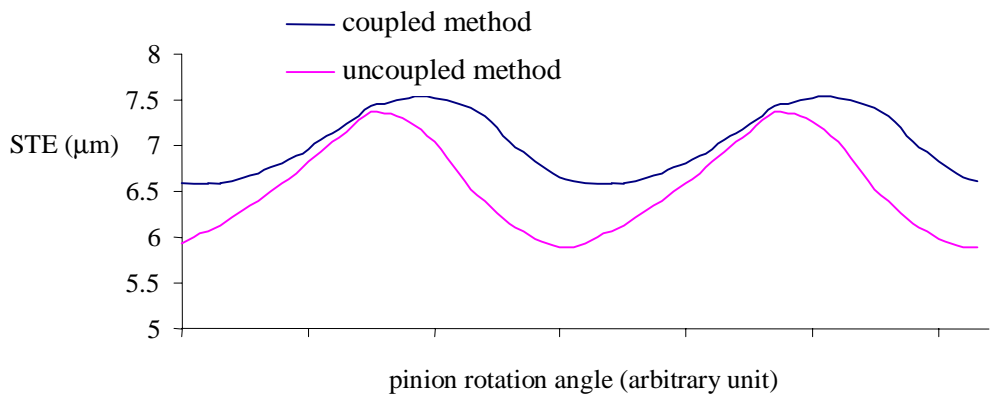

Figure 1: Static transmission error under load

In fact, regarding the load distribution, adaptation of the misalignment (induced in part by bearings deformations) is taken into account with the coupled method. Consequently, average misalignment is equal to –421 µrad with the coupled method instead of –110 µrad which affect the load distribution (see figure 2).

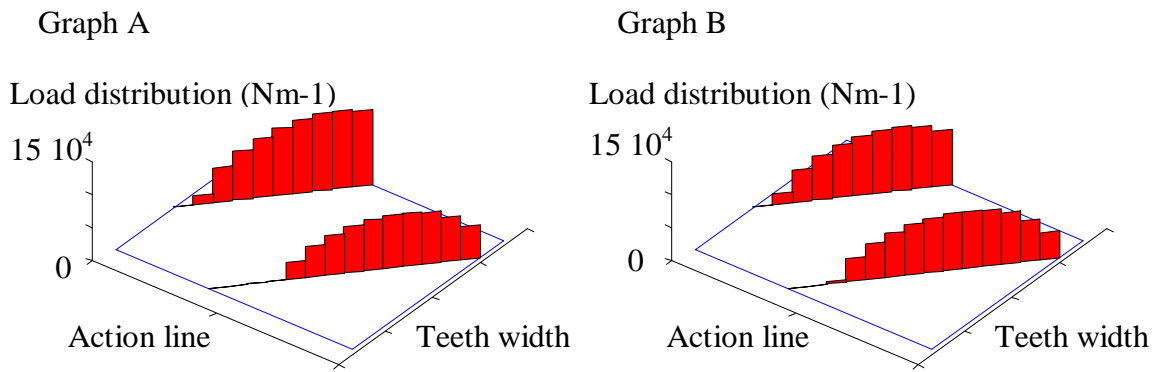

Figure 2: Teeth load distribution : (A) uncoupled and (B) coupled methods

As the contact length is larger for the coupled method, we obtain mesh stiffness equal to 306 N.µm$^{-1}$ instead of 280 N.µm$^{-1}$. Also the bearings stiffness matrices are seriously affected (see table 3).

Table 3: generalized bearings stiffness matrices

| | $F_x$ (daN) | $F_y$ (daN) | $F_z$ (daN) | $M_x$ (mm.daN) | $M_y$ (mm.daN) |
|---|---|---|---|---|---|
| Bearing stiffness with the uncoupled calculus | | | | | |
| x (mm) | 8723 | -2802 | 2122 | -13615 | -38010 |
| y (mm) | -2802 | 3514 | -1040 | 17073 | 13615 |
| z (mm) | 2122 | -1040 | 843 | -6992 | -14349 |
| $\theta_x$ (rad) | -13615 | 17073 | -6992 | 114803 | 91552 |
| $\theta_y$ (rad) | -38010 | 13615 | -14349 | 91552 | 257232 |
| Bearing stiffness with the coupled calculus | | | | | |
| x (mm) | 9648 | -1299 | 2473 | -6965 | -42831 |
| y (mm) | -1299 | 4756 | -532 | 23719 | 6965 |
| z (mm) | 2473 | -532 | 974 | -4153 | -16183 |
| $\theta_x$ (rad) | -6965 | 23719 | -4153 | 145666 | 54369 |
| $\theta_y$ (rad) | -42831 | 6965 | -16183 | 54369 | 283894 |

## 3.3 Dynamic mesh force and vibratory housing response

Dynamic mesh force RMS value versus the output shaft speed is displayed in figure 3. As we can see, some differences appear in levels as in occurrence of critical speeds.

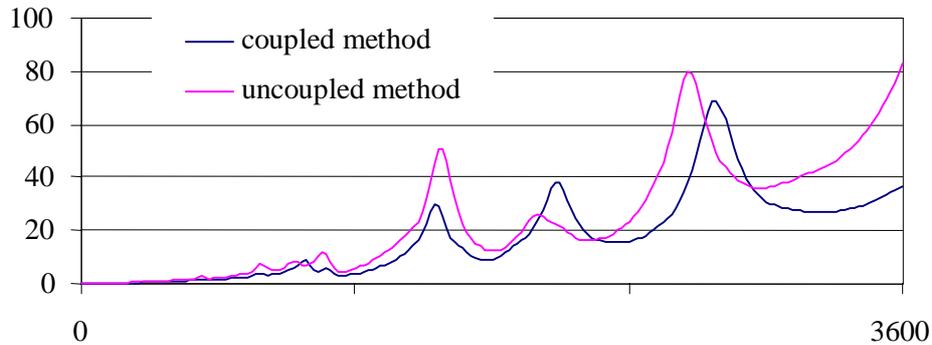

Figure 3: Dynamic mesh force RMS value (N) vs. the output shaft speed (tr/min)

The time and space averaged mean square velocity of the housing is displayed in figure 4. Another time, some discreapencies are obtained. With the assumption of a radiation efficiency equal to one, difference between power radiated noise can be respectively estimated at about 64, 66 and 61 dB for the critical speeds quoted A, B and C in figure 4, instead of 67, 68 and 67 dB with the uncoupled calculation.

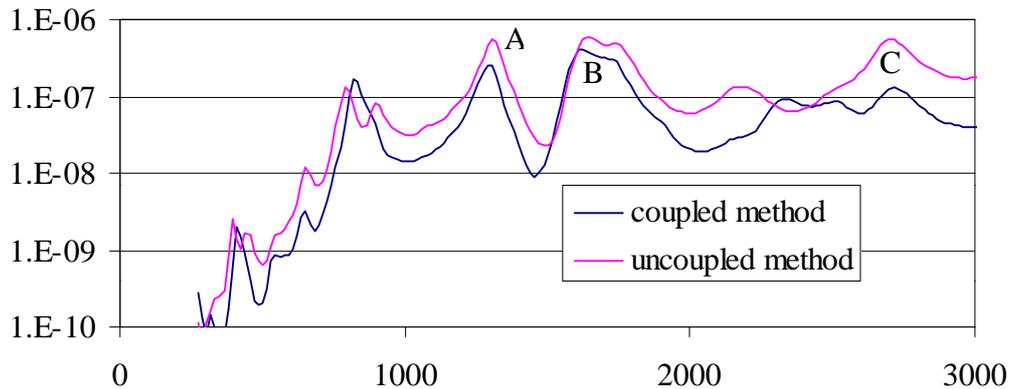

Figure 4: Time- and space-averaged mean square velocity of the housing (m²/s²) vs. the meshing frequency (Hz); A, B, C: critical speeds.

## 4. Conclusion

In this paper we developed a new method for the calculation of STE that takes into account the bearings deformations in a coupling manner. For the chosen gearbox, the interest of this method is demonstrated. In the short future it will be interesting to try different gearboxes system in order to analyze in which configurations the interaction between bearing deformations and mesh equilibrium is not negligible.

## Acknowledgements

The author would like to thank the Region Rhône-Alpes community and the SNR Roulements Company for supporting this work.